\documentclass[aip,jcp,reprint,longbibliography]{revtex4-2}

\usepackage{graphicx}
\usepackage{amsmath}
\usepackage{amssymb}
\usepackage[colorlinks=true,allcolors=blue]{hyperref}
\usepackage[dvipsnames]{xcolor}
\usepackage{afterpage}

\newcommand{\new}[1]{\textcolor{black}{#1}}

\begin{document}

\title{Continuous-time multifarious systems - Part I: equilibrium multifarious self-assembly}

\author{Jakob Metson}
\affiliation{Max Planck Institute for Dynamics and Self-Organization (MPI-DS), 37077 G\"ottingen, Germany}

\author{Saeed Osat}
\email{saeedosat13@gmail.com}
\affiliation{Max Planck Institute for Dynamics and Self-Organization (MPI-DS), 37077 G\"ottingen, Germany}

\author{Ramin Golestanian}
\email{ramin.golestanian@ds.mpg.de}
\affiliation{Max Planck Institute for Dynamics and Self-Organization (MPI-DS), 37077 G\"ottingen, Germany}
\affiliation{Rudolf Peierls Centre for Theoretical Physics, University of Oxford, Oxford OX1 3PU, United Kingdom}

\date{\today}

\begin{abstract}
Multifarious assembly models consider multiple structures assembled from a shared set of components, reflecting the efficient usage of components in biological self-assembly. These models are subject to a high-dimensional parameter space, with only a finite region of parameter space giving reliable self-assembly. Here we use a continuous-time Gillespie simulation method to study multifarious self-assembly and find that the region of parameter space in which reliable self-assembly can be achieved is smaller than what was obtained previously using a discrete-time Monte Carlo simulation method. We explain this discrepancy through a detailed analysis of the stability of assembled structures against chimera formation. We find that our continuous-time simulations of multifarious self-assembly can expose this instability in large systems even at moderate simulation times. In contrast, discrete-time simulations are slow to show this instability, particularly for large system sizes. For the remaining state space we find good agreement between the predictions of continuous- and discrete-time simulations. We present physical arguments that can help us predict the state boundaries in the parameter space, and gain a deeper understanding of multifarious self-assembly.
\end{abstract}


\maketitle

\section{Introduction}
Biological systems operate on a hierarchical principle, wherein smaller building blocks self-assemble to form larger structures, which can in turn assemble into even more complex units~\cite{Kushner1969BR, Whitesides2002S, Zhang2003NB, Stradner2004N, Whitelam2007TJoCP, Whitelam2009SM, Whitelam2015ARPC, Gartner2024PRX, Holmes-Cerfon2025arXiv}. These multi-component systems exhibit behaviors distinct from their single-component counterparts~\cite{Jacobs2013JCP, Jacobs2017BJ, Jacobs2021PRL, Shrinivas2021PNAS, Thewes2023PRL,Parkavousi2025PRL}. A notable feature that biological systems have evolved is their economical utilization of components \cite{Murugan2015NC, Bohlin2023AN}. For instance, a shared pool of amino acids gives rise to a diverse range of protein structures, each fulfilling various functional roles~\cite{Tokuriki2009S, Bornberg-Bauer2013COSB}.
Biological systems have developed innate mechanisms for self-assembly, dictating how components should arrange themselves. In contrast, synthetic self-assembly requires a deliberate design principle to achieve desired target structures \cite{Glotzer2004S, Whitesides2002S, Zhang2004NL, Hormoz2011PNAS, Soto2014PRL,Soto2015PRE,Cademartiri2015NM, Huntley2016PNAS, Nguyen2016PNAS, Zeravcic2017RMP, Klishin2021arXiv, Hagan2021RMP, Goodrich2021PNAS, Klishin2021arXiv, Duffy2021JoAP, Lieu2022TJoCP, McMullen2022N, Gartner2022PNAS, Jhaveri2024PNAS, Evans2024JCP, Metson2025PRR, Duque2024PNAS, Koehler2024PRX, Tyukodi2024arXiv, Benoist2024arXiv, Bassani2024AN, Lieu2025SM, Hubl2025PRL, Hubl2025arXiv}. This pivotal stage of self-assembly, often referred to as the ``design principle'' or ``learning rule'', encompasses the interaction rules between the assembly components. Another bottleneck for self-assembly can arise from frustrations in the interactions that can arise from strong electrostatic interactions that are generically present in polypeptides and polynucleotides \cite{Fazli2005,Mohammadinejad2009}, hampering the accessibility of the desired equilibrium structures. This is particularly relevant for synthetic self-assembly strategies that are based on complementary DNA strand binding, such as DNA-coated colloidal lock-and-key systems \cite{Valignat2005PNAS} and DNA-origami-based systems \cite{Rothemund2006N}. 

The self-assembly of structures from a shared set of components, while appealing, introduces its own set of challenges. One of the primary difficulties arises from the promiscuity of the components. Given that a component can participate in multiple structures, with potentially different neighboring components in each structure, the system becomes susceptible to cross-talk~\cite{Huntley2016PNAS}. This cross-talk between the components has the potential to disrupt the self-assembly process, leading to errors and resulting in the formation of chimeric structures\cite{Murugan2015PNAS, Sartori2020PNAS, Osat2024PRL}. The undesired chimeras form when parts of different structures are stuck together.

\begin{figure}[b]
\begin{center}
\includegraphics[width=\linewidth]{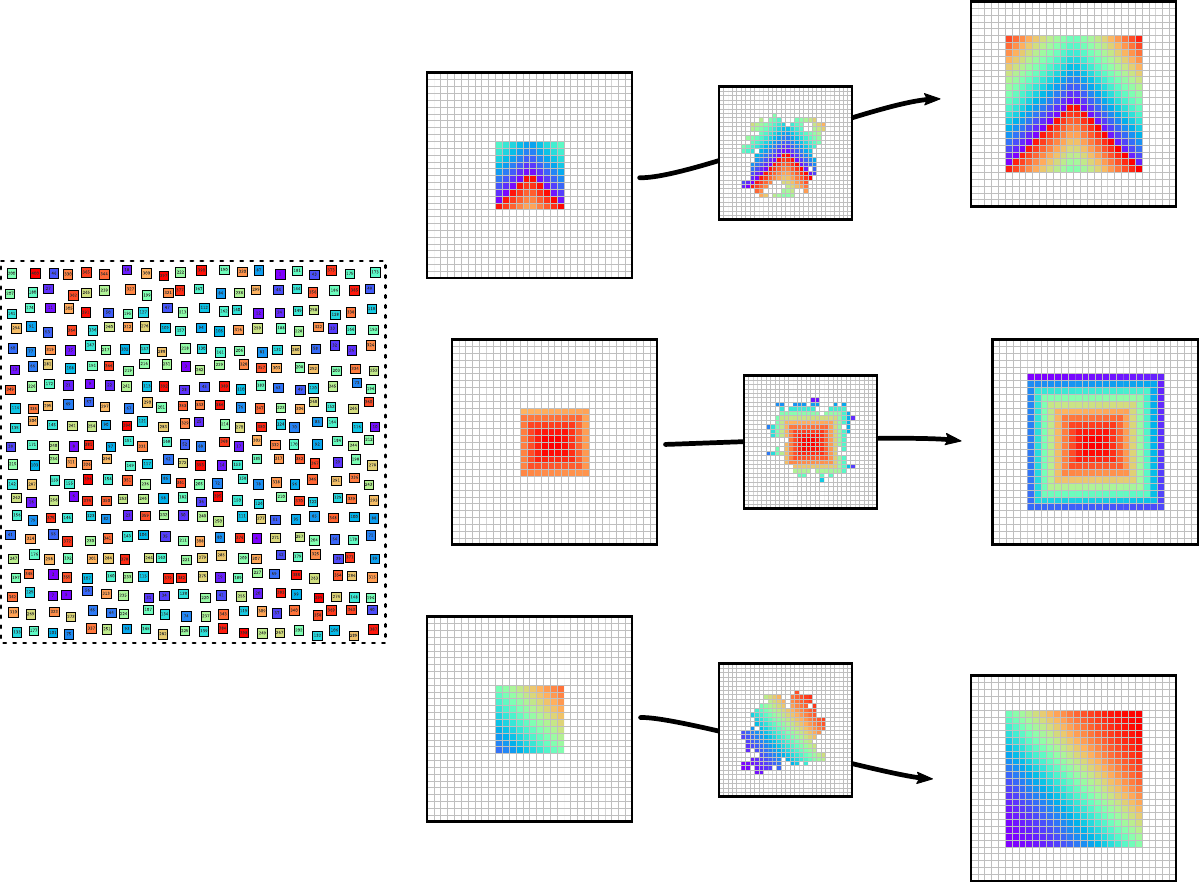}
\caption{An illustration of multifarious self-assembly. Three different structures correspond to particular arrangements of the different tile species. By placing an appropriate seed into the system, the desired structure can be retrieved and assembled. 
\label{fig:Cartoon}}
\end{center}
\end{figure}

Self-assembly suffers from the high dimensionality of the parameter space, i.e., a successful self-assembly design necessitates finely tuned parameters, including temperature, density, and interaction strength among others. Previous research has affirmed that error-free multifarious self-assembly occurs only within a restricted region of this parameter space \cite{Murugan2015PNAS, Sartori2020PNAS, Osat2023NN}. Characterizing this region is crucial for the design of self-assembled systems.

In this paper, we present a continuous-time Gillespie simulation technique for investigating the multifarious self-assembly model. Our study reveals that the region of parameter space conducive to successful self-assembly is narrower than observed in discrete-time Monte Carlo simulations of large systems. We explain the discrepancy between continuous- and discrete-time simulations in the stability of structures against chimera formation and provide analytical calculations of the remaining state boundaries. We verify that for the remaining boundaries, the continuous-time simulations agree with the previously studied discrete-time results. Whilst previous work has demonstrated the functionality of multifarious systems in continuous-time simulations\cite{Zhong2017JSP}, in this work, we explore in detail the parameter ranges that allow for the emergence of stable multifarious self-assembly.

\section{Model}\label{sec:model}

Multifarious self-assembly considers how a shared pool of distinct components can give rise to a myriad of unique macroscopic structures~\cite{Murugan2015PNAS, Sartori2020PNAS, Evans2024N}. This is illustrated in Fig.~\ref{fig:Cartoon}.
The multifarious self-assembly model consists of four primary parts: (i) a shared set of components, (ii) target structures to be assembled from these sets, (iii) interaction rules governing the components, which are dictated by the design principle, and (iv) the selection of parameters within the high-dimensional parameter space of the self-assembly process.

\begin{figure}[tb]
\begin{center}
\includegraphics[width=0.8\linewidth]{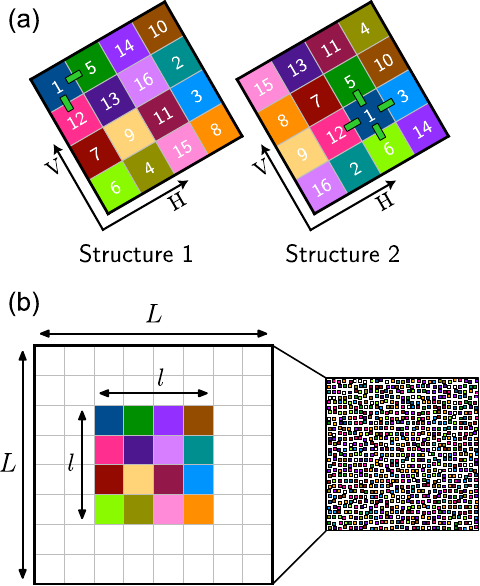}
\caption{An example system of size $L=8$ designed to store two structures of size $l=4$. (a) All the interactions involving tile species 1, following the interaction rule used in this paper. (b) Structure 1 is seen to have been correctly assembled. The reservoir of tiles connected to the system is shown, and the system size $L$ and structure size $l$ are also marked.
\label{fig:Model}}
\end{center}
\end{figure}

The system is designed to store $m$ structures of size $l\times l$ using $M=l\times l$ different tile species arranged on a square lattice. The lattice is size $L\times L$ with hard-wall boundary conditions. 
Each structure contains one tile of each species arranged randomly, as illustrated in Fig.~\ref{fig:Model}(a). This corresponds to a fully heterogeneous setup (every tile in a structure is unique) with zero-sparsity (every tile species is used). 
The interaction rule ensures that for appropriate parameter values and given an appropriate seed, the system is able to autonomously assemble any of the stored structures. In this paper, we use a continuous-time Gillespie algorithm to investigate which parameter values can lead to successful self-assembly, and find differences in the stability boundaries in comparison with the results that use a discrete-time Monte Carlo algorithm. Furthermore, we explore other regions in the state diagram of multifarious systems, namely, dispersion, liquid, and chimera, and present physical arguments to shed light on the differences between the results obtain by the different simulation methods.

We employ the canonical Hebbian interaction rule used often in the multifarious-self assembly literature \cite{Murugan2015PNAS, Sartori2020PNAS, Osat2023NN, Teixeira2024PNAS}, which is based on the Hopfield model\cite{Hopfield1982PNAS}. First, we define ${\cal I}^{\mathrm{r}}$ as the set of all specific reciprocal interactions that are required to ensure that the desired $m$ structures are stored in the system as minimum-energy configurations. This set is a union of smaller sets that each contains the specific interactions imposed by one of the target structures, ${\cal I}^{\mathrm{r}} \equiv 
I^{\mathrm{r}} (S^1) \cup I^{\mathrm{r}} (S^2) \cup \dots \cup I^{\mathrm{r}} (S^m)$, where $S^\ell$ denotes the $\ell$\textsuperscript{th} structure stored in the system.
The smaller sets can be explicitly defined as follows
\begin{align}
    I^{\mathrm{r}} (S^{\ell}) = \bigcup_{\substack{ {\left \langle \alpha,\beta \right\rangle } }} \; 
    S^{\ell}_{\alpha}  \square S^{\ell}_{\beta},
    \label {eq:interactions}
\end{align}
where $\alpha$ and $\beta$ are lattice coordinates $(i,j)$ running over nearest neighbors and $\square \in \{ \diagdown, \diagup \}$ represents the directions of the specific interactions, horizontal or vertical as shown in Fig.~\ref{fig:Model}(a). Equation~\eqref{eq:interactions} states that a specific interaction belongs to the set of reciprocal interactions $I^{\mathrm{r}} (S^{\ell})$ if at least one target structure favors that interaction. 
The interaction matrix is then 
\begin{align}
  U_{A \square B}^{\mathrm{r}} = \begin{cases}
    -\varepsilon, & \text{if } A \square B \in  {\cal I}^{\mathrm{r}}, \\
    0, & \text{otherwise}.
  \end{cases}
  \label{eqn:U}
\end{align}
A key advantage of this design scheme is that we have a single parameter $\varepsilon$ controlling the interaction strength, greatly reducing the dimensionality of the parameter space.

Each of the lattice sites in the system is either occupied by the solvent (empty) or occupied by a tile of a particular species. We consider each lattice site to be connected to a shared reservoir of solvent and tiles. The concentration $c_i$ of tile species $i$ in the reservoir is obtained by normalizing the number of species $i$ tiles in the reservoir $N_i$ by the number of solvent tiles in the reservoir $N_0$,
\begin{equation}
    c_i = \frac{N_i}{N_0}.
\end{equation}
From the concentrations, we obtain the chemical potential of tile species $i$
\begin{equation}
    \mu_i = k_{\rm B} T \log c_i.
\end{equation}
We assume that the reservoir is very large, such that the chemical potential of each species can be considered constant throughout the simulation. Each lattice site is connected to the same reservoir, which corresponds to the assumption of fast diffusion in the reservoir, meaning there are no gradients in chemical potential across the lattice. By definition, the chemical potential of the solvent is zero. Within each simulation we set the chemical potential of the tile species to the same value $\mu$. We set $k_{\rm B} T = 1$, and absorb the temperature into the now dimensionless parameters $(-\mu,\varepsilon)$.

Once the structures have been defined and the interactions calculated, a chosen seed is placed into the lattice and the system is then simulated. For clarity of presentation, we use a specific coloring scheme in which tiles belonging to the same target structure will be colored similarly; see Appendix \ref{app:Colouring} for more details. In this paper we use two different algorithms to simulate multifarious self-assembly: a discrete-time Monte Carlo algorithm and a continuous-time Gillespie algorithm. 
Both algorithms simulate reactions where tiles are exchanged between the lattice and the reservoir.
The algorithms are explained in detail in Appendix~\ref{app:algorithms} and a flowchart detailing the steps of the two simulation methods is shown in Fig.~\ref{fig:Algorithms}.

\begin{figure}[tb]
\begin{center}
\includegraphics[width=0.99\linewidth]{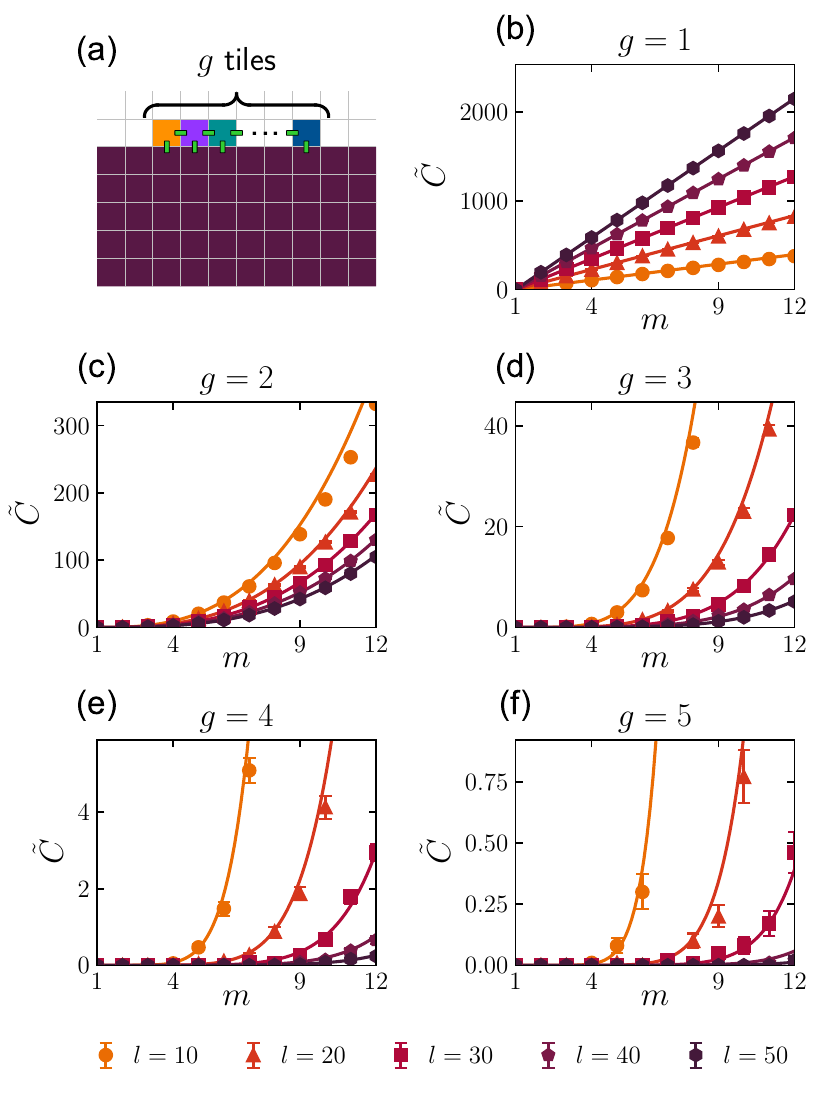}
\caption{(a) An illustration of a fully connected strip of $g$ tiles, attached to the boundary of an assembled structure. Expected number of chimeric strip configurations $\tilde{C}$ of length (b) $g=1$, (c) $g=2$, (d) $g=3$, (e) $g=4$, (f) $g=5$, plotted as a function of $m$, the number of patterns stored in the system for different pattern sizes $l\times l$. The analytic expression Eq.~\eqref{eq:number_chim_configs} is plotted as the solid lines and agrees well with the number of configurations measured in simulations shown by the data points, using 100 simulations per data point.
\label{fig:CountingChimConfigs}}
\end{center}
\end{figure}

\begin{figure*}[tb]
\begin{center}
\includegraphics[width=1.0\linewidth]{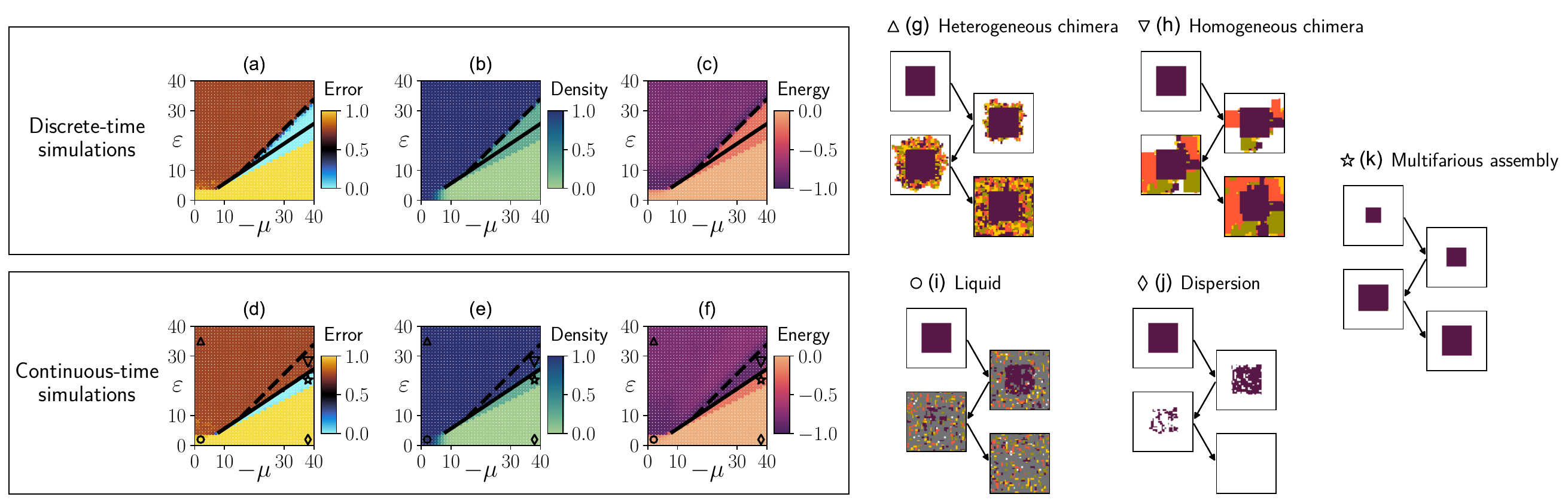}
\caption{Plots of error [(a) and (d)], density [(b) and (e)], and energy [(c) and (f)], measured using discrete- and continuous-time simulations, respectively. Plotted in $(-\mu, \varepsilon)$ state space for systems with $m=4$ stored structures. 
The dashed black line marks the multifarious assembly-chimera boundary in discrete-time Monte Carlo simulations, and the solid black line marks the same boundary in continuous-time Gillespie simulations.
To start the simulations a $20\times 20$ pattern is placed in a $40\times 40$ system. From the three quantities plotted we can distinguish the different states (heterogeneous chimera, homogeneous chimera, liquid, dispersion and multifarious assembly). (g)-(k) Snapshots of the different states using the continuous-time Gillespie algorithm. Parameter values are given by the corresponding markers in the state diagram. Snapshots using discrete-time simulations look similar and can be found in previous works\cite{Murugan2015PNAS, Sartori2020PNAS, Osat2023NN}. Snapshots are colored by structure, as described in Appendix~\ref{app:Colouring}.
\label{fig:phase_dig_gillespie_vs_mc}}
\end{center}
\end{figure*}

\section{Analyzing the stability of structures}\label{sec:g-tile_strip}
A necessary condition for successful self-assembly is that the structure is stable. In this section, we outline a method to find the region of parameter space that will give rise to stable structures. To analyze the stability of a self-assembled structure we consider two processes: adding a strip of $g$ tiles to a flat boundary of a structure, and removing such a strip. The strip setup is depicted in Fig.~\ref{fig:CountingChimConfigs}(a). The ratio of the rate of forming such a strip ($K_{\rm f}$) to the rate of removing it ($K_{\rm r}$) is calculated in Appendix~\ref{app:KfKr}. We obtain
\begin{equation}\label{eq:KfKr_maintext}
    \frac{K_{\rm f}}{K_{\rm r}} = e^{\mu} e^{2\varepsilon} \cdot \frac{(g-1) + e^{-\varepsilon/2}}{(g-1) + e^{\varepsilon/2}}.
\end{equation}
Laying a strip of length $g$ becomes favorable when
\begin{equation}\label{eq:laying_favourable}
    e^{\mu} e^{2\varepsilon} \cdot \frac{(g-1) + e^{-\varepsilon/2}}{(g-1) + e^{\varepsilon/2}} > 1.
\end{equation}
This result can be used to explain the stability of assembled structures against dissolving into dispersion, as shown in Section~\ref{sec:MA-D_boundary}, and to derive the critical seed size for assembly, as shown in Appendix~\ref{app:critical_seed_size}. Furthermore, to predict the stability of assembled structures against forming chimeras, we can consider the attachment of a strip of $g$ tiles to the boundary of an assembled structure. If this process is favorable, then the system will continue to build off of the assembled structure leading to the formation of a chimera. As shown in Fig.~\ref{fig:CountingChimConfigs}(a), we assume that each tile in the strip makes bonds to all of the neighbors it has. 
For the nucleation of chimeric structures onto the boundary of an assembled structure we may not always have the required interactions to form a fully bonded strip. To quantify this, we calculate $\tilde{C}$, the expected number of fully-bonded chimeric strips of length $g$ that could be added to a structure of size $l\times l$ in a system with $m$ stored  structures. The derivation is given in Appendix~\ref{app:CountingChimConfigs} and results in
\begin{equation}\label{eq:number_chim_configs}
    \tilde{C}(g,m,l) = 4(l+1-g)(m-1)^g \left(1-\frac{1}{l}\right)^{2g-1} \left(\frac{m}{l^2-1}\right)^{g-1}.
\end{equation}
This expression is plotted against results measured in simulations in Fig.~\ref{fig:CountingChimConfigs}(b)-(f), and shows good agreement. In Section~\ref{sec:Ch-MA_boundary}, we use Eqs.~\eqref{eq:laying_favourable}~and~\eqref{eq:number_chim_configs} to rationalize the stability of an assembled structure against chimera formation.


\section{State diagram}
We use the error, density, and energy to characterize the state of the system. The error reflects how far away the current configuration of tiles in the lattice is from an assembled structure. The density is the number of tiles in the lattice normalized to the lattice size, and the energy represents the number of bonds being made in the lattice. In Appendix~\ref{app:orderparams}, we provide precise definitions of these three quantities.

The state diagram of discrete-time Monte Carlo simulations of multifarious self-assembly\cite{Murugan2015PNAS, Sartori2020PNAS} is shown in Figs.~\ref{fig:phase_dig_gillespie_vs_mc}(a)-(c). In continuous-time Gillespie simulations, shown in Figs.~\ref{fig:phase_dig_gillespie_vs_mc}(d)-(f), we find the same states as in discrete-time simulations, namely
\begin{itemize}
    \item heterogeneous chimera: high error, high density, low energy,
    \item homogeneous chimera: high error, high density, very low energy,
    \item multifarious-self assembly: low error, medium density, low energy,
    \item liquid: high error, high density, high energy,
    \item dispersion: high error, low density, high energy.
\end{itemize}

\begin{figure*}[t]
\begin{center}
\includegraphics[width=0.8\linewidth]{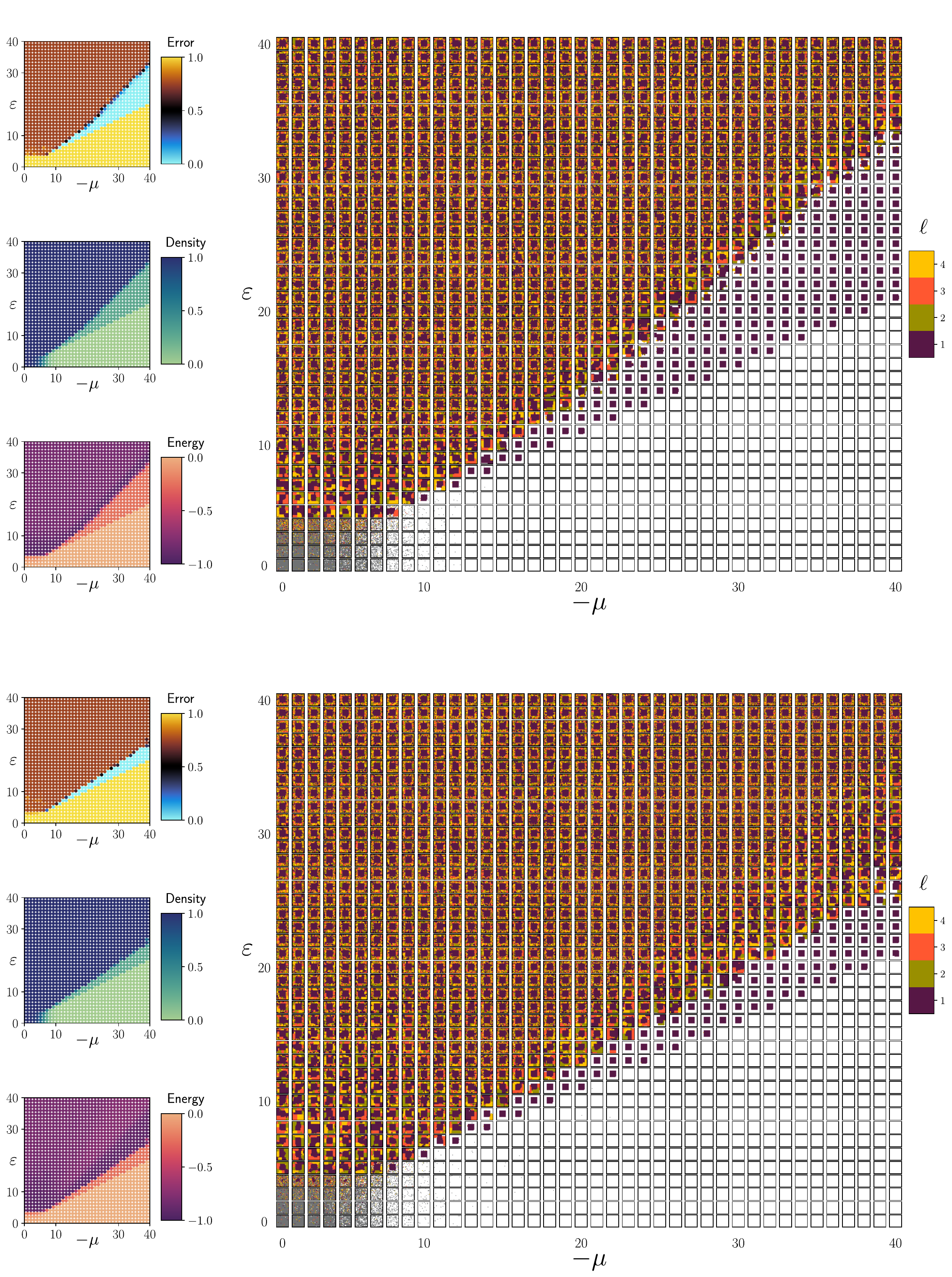}
\caption{State diagram with snapshots. Top: Discrete-time Monte Carlo simulations. Bottom: Continuous-time Gillespie simulations. Snapshots are colored by structure, as described in Appendix~\ref{app:Colouring}. For each value of $(-\mu,\varepsilon)$ a system with $m=4$ stored $20\times 20$ structures is simulated. To start the simulations a $20\times 20$ pattern is placed in the $40\times 40$ system.
\label{fig:phase_dig_snapshots}}
\end{center}
\end{figure*}

\afterpage{\clearpage}

Snapshots of these different states are shown in Figs.~\ref{fig:phase_dig_gillespie_vs_mc}(g)-(k). State diagrams with snapshots are shown in Fig.~\ref{fig:phase_dig_snapshots}.
The chimera state is divided into two sub-states. Homogeneous chimeras are characterized by very low energies, corresponding to large domain sizes. Heterogeneous chimeras have smaller domain sizes and slightly higher energies, and occur when $\varepsilon$ is large.
As shown in Fig.~\ref{fig:phase_dig_gillespie_vs_mc}(d)-(f), the topology of the state diagram is the same in continuous-time as in discrete-time simulations. We have the following boundaries: liquid-dispersion, dispersion-assembly, chimera-assembly and chimera-liquid. Although we find the same topology, we observe in Fig.~\ref{fig:phase_dig_gillespie_vs_mc} that the region corresponding to stable multifarious self-assembly is smaller when using continuous-time Gillespie simulations. In particular, the assembly-chimera boundary shifts downwards. We can rationalize this using the framework outlined in Section \ref{sec:g-tile_strip}, which is done in the following subsection. We also provide calculations of the other three boundaries in the following subsections. Only the assembly-chimera boundary is changed when using continuous-time instead of discrete-time simulations. The other three boundaries (dispersion-assembly, liquid-dispersion and chimera-liquid) remain consistent between the two simulation methods.
For the dispersion-assembly and assembly-chimera boundaries, we focus on deriving the scalings of the boundaries. We use the $\sim$ and $\gtrsim$ symbols to indicate that constant offset terms can also be present \cite{Sartori2020PNAS}.

\subsection{Assembly-chimera boundary}\label{sec:Ch-MA_boundary}
\begin{figure}[tb]
\centering
\includegraphics[width=0.99\linewidth]{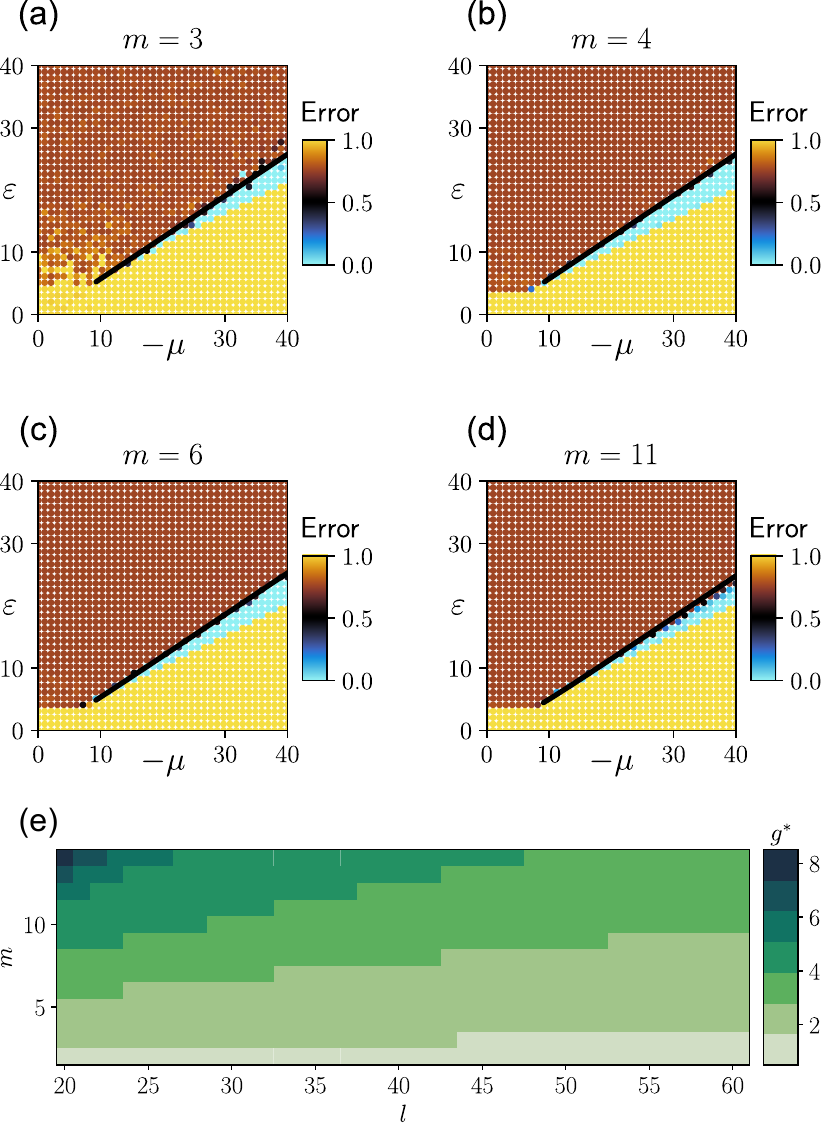}
\caption{
Upon increasing $m$, the number of structures stored in the system, we observe a shift in the assembly-chimera boundary, in accordance with the predicted boundary scalings derived in Section~\ref{sec:Ch-MA_boundary} shown by the solid black lines here.
(a) Discrete-time Monte Carlo simulations with system size $L=16$ and structure size $l=8$.
(b)-(d) Continuous-time Gillespie simulations with system size $L=40$ and structure size $l=20$.
(e) Values of $g^*$ for different structure sizes $l$ and numbers of stored structures $m$.
\label{fig:shifting_Ch-MA_boundary}}
\end{figure}

In continuous-time Gillespie simulations, the assembly-chimera boundary is shifted down from the $\varepsilon \sim -\mu$ boundary seen in discrete-time simulations. 
Part of the assembly region which seemed stable in discrete-time simulations turns out to be unstable to chimera formation. Assembled structures are unstable to chimera formation if the system is able to add tiles to the boundary of the assembled structure, to form chimeric structures.
In Section \ref{sec:g-tile_strip} we derived the condition for laying a strip of length $g$ being more favorable than removing such a strip, giving Eq.~\eqref{eq:laying_favourable}. Rearranging this condition we obtain
\begin{equation}\label{eq:eps_favourable}
    2\varepsilon + \log\left(\frac{(g-1) + e^{-\varepsilon/2}}{(g-1) + e^{\varepsilon/2}}\right) > -\mu.
\end{equation}
Provided a valid configuration exists, structures with side lengths greater than $g$ will be unstable to chimera formation.
In this region, the lifetime of structures is similar to the assembly time, and so we do not consider these structures to be stable.
However, we may not always have a valid chimeric configuration. In Section \ref{sec:g-tile_strip} we also give the expected number of chimeric strip configurations (Eq.~\eqref{eq:number_chim_configs}). Putting these two concepts together, we can deduce that the assembly-chimera boundary in a given system is
\begin{equation}
    2\varepsilon + \log\left(\frac{g^*-1 + e^{-\varepsilon/2}}{g^*-1 + e^{\varepsilon/2}}\right) \sim -\mu.
\end{equation}
where $g^*$ is the largest value of $g$ for which
\begin{equation}
\tilde{C} \gtrsim 1.
\end{equation}
The expression for $\tilde{C}$ is given by Eq.~\eqref{eq:number_chim_configs}, and a plot of $g^*$ as a function of $l$ and $m$ is shown in Fig.~\ref{fig:shifting_Ch-MA_boundary}(e).
Since larger strips are more favorable to form, the instability is controlled by the largest possible value of $g$ for a given set of interactions.
In the limits of small and large $(\varepsilon,\mu)$ we find two different linear scaling relationships. 
For small $\varepsilon$ and $\mu$ we obtain
\begin{equation}\label{eq:MACh_bound_smalleps}
    \varepsilon \sim -\frac{g^*}{2g^*-1}\mu,
\end{equation}
whilst for large $\varepsilon$ and $\mu$ (and $g^*>1$) we obtain
\begin{align}\label{eq:MACh_bound_largeeps}
    \varepsilon \sim -\frac{2}{3}\mu - \frac{2}{3}\log{(g^*-1)}.
\end{align}
Figures \ref{fig:shifting_Ch-MA_boundary}(b)-(d) show how the assembly-chimera boundary shifts in continuous-time Gillespie simulations as we change $m$, the number of patterns stored in the system. As $m$ increases, the assembly-chimera boundary shifts down as more of the $(-\mu,\varepsilon)$ state space becomes unstable to chimera formation. This is consistent with the concept that increasing the number of stored structures increases the cross-talk between structures, leading to reduced stability. 

For $\varepsilon<-\mu$, tiles need more than one bonding neighbor to be favorably added. Therefore the homogeneous chimeras with large domain sizes are the first to emerge when crossing from the assembly region to the chimera region. For $\varepsilon>-\mu$, tiles can be added with only one binding neighbor, leading to many possible chimera nucleation events and the resulting higher-energy heterogeneous chimeras with small domain sizes. 

Since the specific interactions in each system are derived from random configurations, the actual number of chimeric strips possible in a given system is not always exactly the expected number. This can change the stability for an individual system, particularly when $g^*$ is close to being between two different values. However the method developed here correctly predicts the chimera-assembly boundary, which is defined by the average system behavior.

In the continuous-time Gillespie simulations, once one tile of a strip has been placed, the reactions which further grow the strip are likely to be picked due to the increased rate of these reactions relative to other processes in the system.
In contrast, in discrete-time simulations reactions are proposed randomly. The average number of Monte Carlo steps needed to sweep every reaction scales $\propto l^4$, growing rapidly as the structure size is increased \cite{Osat2024PRL}. Furthermore, for $g>1$, the instability requires multiple consecutive reactions to occur. This greatly amplifies the timescales, since following the nucleation of a strip, conflicting reactions may be executed before the strip can grow further. For example, the reaction to remove the first tile of a strip will always be accepted for $\varepsilon<-\mu$.
This clarifies why in discrete-time Monte Carlo simulations of medium- to large-sized structures, chimeras formed by nucleation of a new structure out of the central structure are only observed for a small band of $(-\mu, \varepsilon)$ values close to $\varepsilon\sim-\mu$. Moreover, it explains why a significant portion of the region we find to be unstable in continuous-time Gillespie simulations is stable in the discrete-time simulations.
As shown in Fig.~\ref{fig:shifting_Ch-MA_boundary}(a), the boundaries predicted in this section are indeed observed in discrete-time simulations, but only for small systems where the required simulation time for instability is much shorter.

\subsection{Dispersion-assembly boundary}\label{sec:MA-D_boundary}
Using the $g$-strip construction, we can also analyze the stability of structures against dissolving into dispersion. Taking the $g\to \infty$ limit of Eq.~\eqref{eq:laying_favourable} we get
\begin{equation}
    \varepsilon > -\frac{1}{2}\mu.
\end{equation}
For $\varepsilon < -\mu/2$ it is always more favorable to remove a strip than to lay it, even for infinitely long strips, meaning that all structures will dissolve into dispersion. Since it is always possible to remove an existing strip, we do not need to consider the expected number of configurations. As can be seen in Fig.~\ref{fig:phase_dig_gillespie_vs_mc}, this transition is the same for discrete- and continuous-time simulations. This transition can also be understood by considering peeling off the corners of a structure. 
For low $\varepsilon$, as $|\mu|$ is reduced, the system transitions from dispersion (low density) to liquid (high density).

\subsection{Liquid-dispersion boundary}\label{sec:L-D_boundary}

\begin{figure}[tb]
\includegraphics[width=0.9\linewidth]{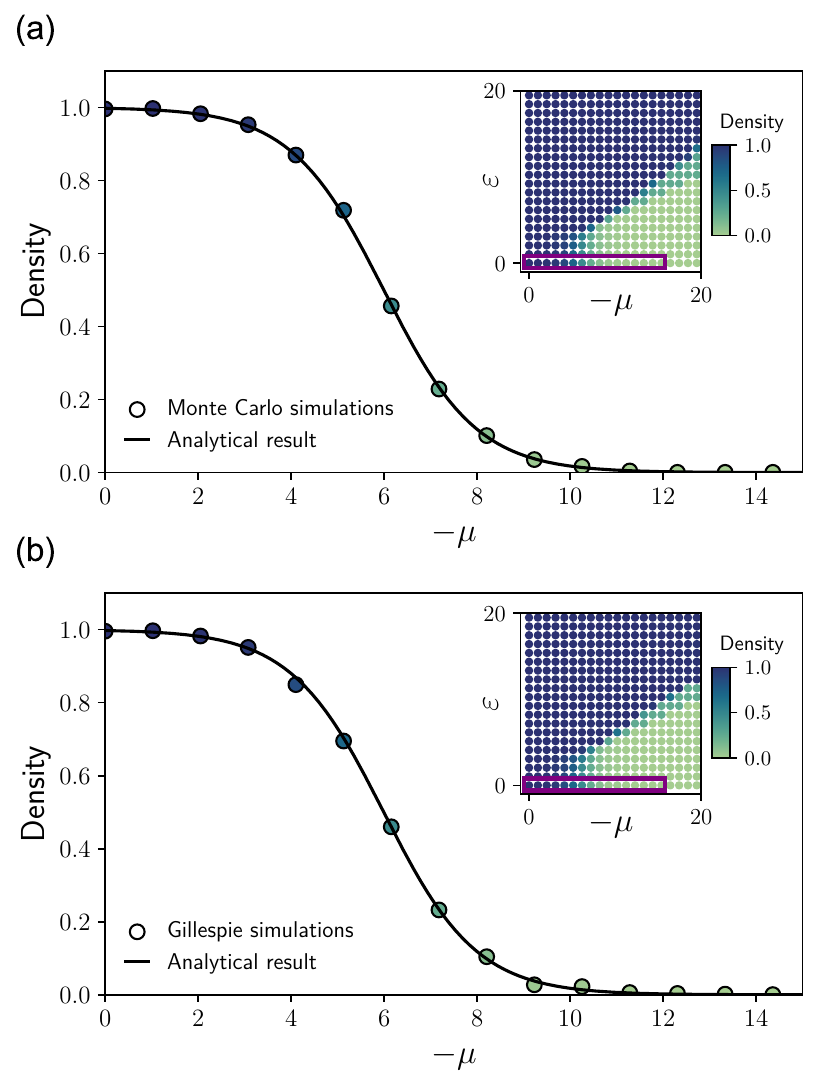}
\caption{Liquid-dispersion transition. A cut along the line $\varepsilon=0$ is taken, as shown by the purple box on the inset state diagrams. The density along this cut is plotted as a function of $\mu$, for (a) discrete-time Monte Carlo and (b) continuous-time Gillespie simulations. The density transition is quantitatively the same using the two different algorithms. The solid line plots the theoretical prediction Eq.~\eqref{eq:density_mu}, showing a near-exact agreement. The data plotted here is the same as in Fig.~\ref{fig:phase_dig_gillespie_vs_mc}.
\label{fig:L-DS_transition}}
\end{figure}

The liquid-dispersion transition is characterized by the density. In the liquid state (small negative $\mu$) the density is close to one, whilst in the dispersion state (large negative $\mu$) the system is dilute. At $\varepsilon=0$, only the chemical potential of the tiles will affect the rate of different reactions, and so we can predict the liquid-dispersion transition by considering the concentrations of tiles in the reservoir. The derivation given in Appendix~\ref{app:L-D_transition} leads to the following expression for density variation as a function of $-\mu$ along the cut $\varepsilon=0$
\begin{equation}\label{eq:density_mu}
\rho(-\mu) = \frac{M}{M+e^{-\mu}}.
\end{equation}
This theoretical argument correctly predicts the density variation across the transition, as shown in Fig. \ref{fig:L-DS_transition}.

\subsection{Chimera-liquid boundary}\label{sec:L-Ch_boundary}

\begin{figure}[tb]
\includegraphics[width=0.9\linewidth]{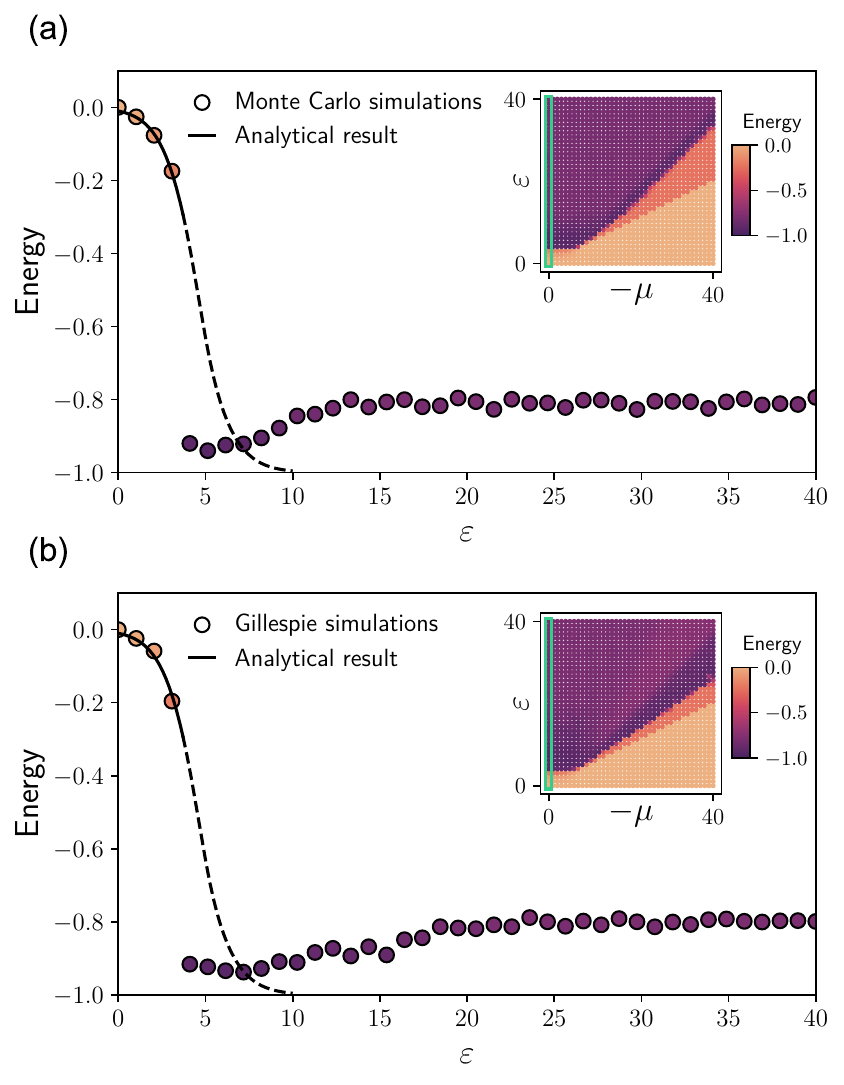}
\caption{Chimera-liquid transition. A cut along the line $\mu=0$ is taken, as shown by the green box on the inset state diagrams. The normalized energy along this cut is plotted as a function of $\varepsilon$, for (a) discrete-time Monte Carlo and (b) continuous-time Gillespie simulations. The transition is similar using the two different algorithms. 
The solid line plots the theoretical mean-field result Eq.~\eqref{eq:energy_epsilon}, which works well for small $\varepsilon$ where the turnover of tiles is high. The data plotted here is the same as in Fig.~\ref{fig:phase_dig_gillespie_vs_mc}.
\label{fig:LChTransition}}
\end{figure}

The chimera-liquid transition is characterized by the energy, or number of bonds, in the system. In the liquid state (small $\varepsilon$) most tiles in the system are not interacting, leading to very few bonds in the system. At higher $\varepsilon$ we reach the chimera state, characterized by a large number of bonds being made in the system.

In Appendix~\ref{app:L-Ch_transition}, we present a mean-field calculation for the average number of bonds per tile $\nu$, leading to the following expression for the normalized energy
\begin{equation}\label{eq:energy_epsilon}
E = -\frac{1}{4}\frac{4P_4 e^{4\varepsilon} + 3P_3 e^{3\varepsilon} + 2P_2 e^{2\varepsilon} + P_1 e^{\varepsilon}}{P_4 e^{4\varepsilon} + P_3 e^{3\varepsilon} + P_2 e^{2\varepsilon} + P_1 e^{\varepsilon} + P_0},
\end{equation}
where the expressions for $P_v$ are given in Appendix~\ref{app:L-Ch_transition}. As can be seen in Fig.~\ref{fig:LChTransition}, this theory works well for small $\varepsilon$ in the liquid state, where the turnover of tiles is high, as explained in Appendix~\ref{app:L-Ch_transition}. For higher $\varepsilon$, the system becomes trapped due to the high energy of bonds, and becomes unable to rearrange itself to lower energy configurations. This leads to the non-monotonic behavior observed in Fig.~\ref{fig:LChTransition} \cite{Sartori2020PNAS}.

\section{Conclusions}
In this paper, we explore the behavior of multifarious self-assembly systems simulated in continuous time. 
\new{For the majority of state space, the two simulation methods are in agreement. Furthermore, the topology of the state diagram remains unchanged. This is a useful validation of the discrete-time Monte Carlo dynamics used to date, particular for these kinds of systems, where the dynamics on observable timescales often evolves through metastable states far from the true equilibrium configuration.}
However, we do find that a region of parameter space giving stable structures in discrete-time Monte Carlo simulations is unstable to chimera formation in continuous-time Gillespie simulations. The extent of this region depends on the size and number of stored structures. By considering the nucleation of a strip of tiles to the boundary of an assembled structure, we explain and predict the scaling of the chimera-assembly boundary, which determines the range of the discrepant region. 
\new{In particular, we find that the scaling of the boundary between stable multifarious-assembly and chimeras changes from $\varepsilon\sim-\mu$ in discrete-time Monte Carlo simulations to $\varepsilon\sim-\frac{2}{3}\mu$ in continuous-time Gillespie simulations.}
\new{Which simulation algorithm is more physically correct depends on the underlying system being studied. One must additionally consider the other abstractions used in these models.}
We also make calculations of the other boundaries in the state diagram, with exact calculations of the assembly-dispersion and liquid-dispersion boundaries, and an approximate calculation for the chimera-liquid boundary. These calculations can be applied to both the discrete-time and continuous-time pictures, as verified by our simulations.

Non-equilibrium versions of multifarious self-assembly have also been of great interest recently \cite{Bisker2018PNAS, Osat2023NN, Faran2023JPCB, Osat2024PRL, Faran2025JCIM}. Continuous-time simulations of a non-reciprocal multifarious self-organization system are addressed in a following work \cite{Metson2025arXiv-b}.

\begin{acknowledgments}
We acknowledge support from the Max Planck School Matter to Life and the MaxSynBio Consortium which are jointly funded by the Federal Ministry of Education and Research (BMBF) of Germany and the Max Planck Society.\end{acknowledgments}

\bibliography{refs, refs_JM,Golestanian}


\clearpage
\appendix

\section{Coloring lattice snapshots}\label{app:Colouring}
\begin{figure}[tb]
\begin{center}
\includegraphics[width=\linewidth]{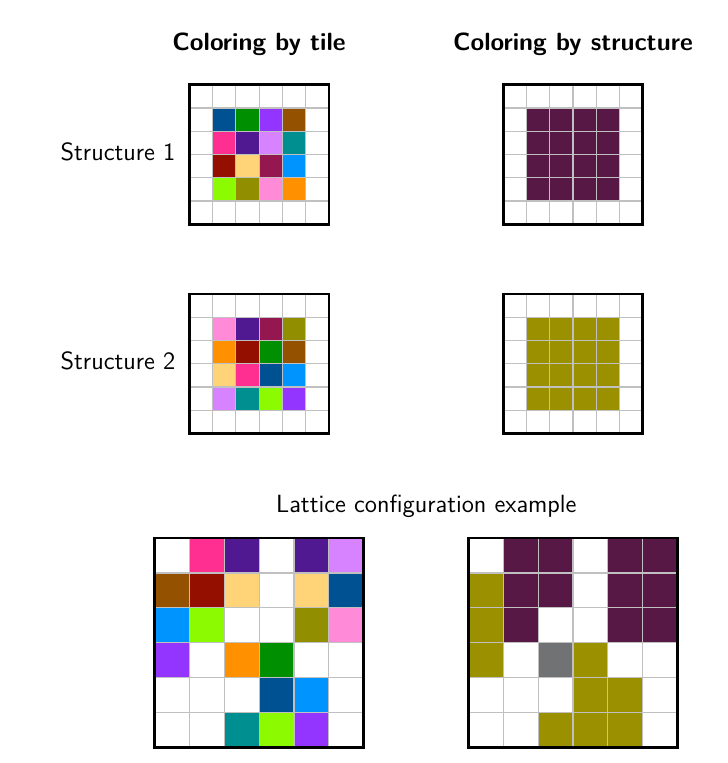}
\caption{An illustration of coloring by tile species and coloring by structure. For clarity we generally color snapshots by structure. White corresponds to empty lattice sites. Tiles which are present in the lattice with no specific interactions are colored gray. All other tiles present are colored based on the specific interactions they make within each structure.
\label{fig:colouring}}
\end{center}
\end{figure}

To display snapshots of the system lattice at a given time in a simulation, one could simply assign a unique color to each tile species, as done in Fig.~\ref{fig:Cartoon}. However this quickly becomes hard to visually interpret, especially for structures defined as random arrangements of the different tile species. Therefore we color snapshots based on the different structures, or parts of structures, assembled. First we assign a unique color to each of the $m$ structures stored in the system.
Empty lattice sites are colored white. Tiles which are present in the lattice with no specific interactions are colored gray. For each remaining tile, we count through its neighbors in the lattice and assign it the color of the structure for which it has the most neighbors currently in the lattice.
An illustration of the coloring by tile and by structure schemes is shown in Fig.~\ref{fig:colouring}.

\section{Simulation algorithms}\label{app:algorithms}

In this paper we use both a discrete-time Monte Carlo\cite{Newman1999} algorithm and a continuous-time Gillespie\cite{Gillespie1976JCP, Gillespie1977JPC, Newman1999} algorithm to simulate the multifarious self-assembly model. In this Appendix section we describe in detail the two algorithms we use.

The configuration of the lattice at any given time can be captured by the configuration variable $\sigma_{\alpha}$, with $\alpha$ representing the lattice coordinates. $\sigma_{\alpha}=1, 2, \dots ,M$ represents different tile species, while $\sigma_{\alpha}=0$ stands for an empty lattice site.

\subsection{Discrete-time Monte Carlo simulation}
At each step in discrete-time Metropolis Monte Carlo simulations of the system, a random reaction is proposed. For example, changing the tile $\sigma_{i,j}$ located at lattice site $\alpha=(i,j)$ to the tile $\sigma'$.
This move is accepted with probability
\begin{equation}
p_\mathrm{acc} = \min[1, e^{-\Delta \mathcal{H}}],
\end{equation}
where
\begin{equation}
\mathcal{H} =  \sum_{\langle \alpha, \beta \rangle} U^{\rm r}_{\sigma_\alpha \square \sigma_\beta} 
 - \mu \sum_{\alpha} (1-\delta_{0,\sigma_\alpha}).
\end{equation}

If accepted, the system is updated according to the chosen reaction. Otherwise, the system remains unchanged.

\begin{figure*}[tb]
\begin{center}
\includegraphics[width=\linewidth]{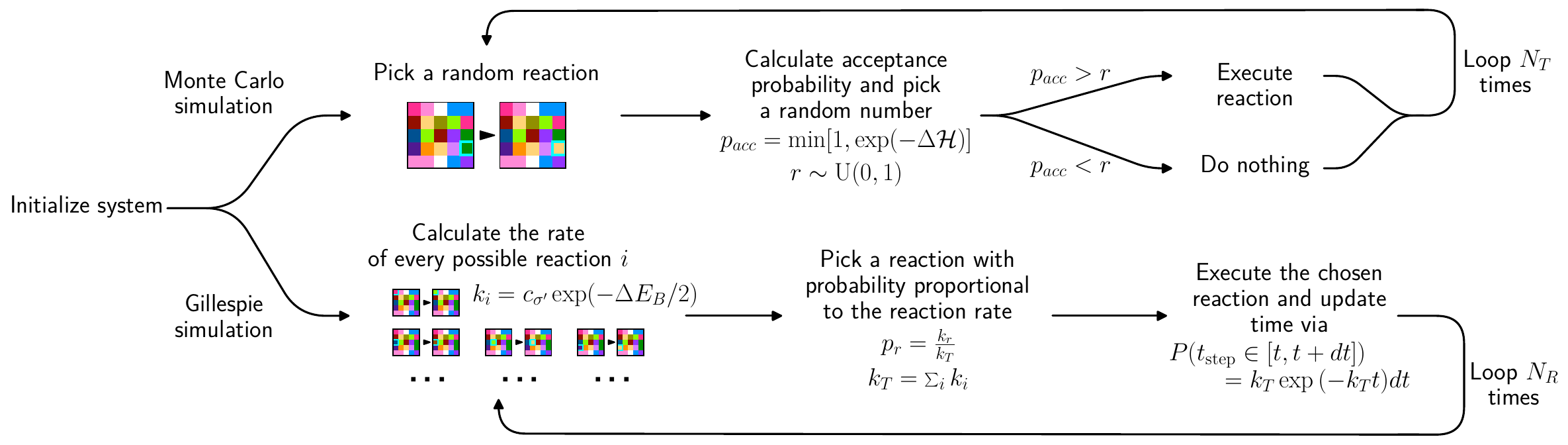}
\caption{Flowchart of the discrete-time Monte Carlo and continuous-time Gillespie simulations algorithms used in this paper to simulate the multifarious self-assembly model. $\mathrm{U}(0,1)$ denotes a uniform distribution over the interval $[0,1]$.
\label{fig:Algorithms}}
\end{center}
\end{figure*}

\subsection{Continuous-time Gillespie simulation}
To simulate this system using the continuous-time Gillespie algorithm, at each step we calculate the rate of all possible reactions in the system. The rate of changing the tile $\sigma_{i,j}$ located at lattice site $(i,j)$ to the tile $\sigma'$ is
\begin{equation}
    k_{\sigma_{i,j} \rightarrow \sigma'} = c_{\sigma'} e^{-\Delta E_B/2},
\end{equation}
where $c_{\sigma'}$ is the concentration of species $\sigma'$ in the reservoir, as explained in Section~\ref{sec:model}. The change in bond energy is
\begin{multline}
\Delta E_B = 
U^{\rm r}_{\sigma_{i-1,j} \diagup \sigma'}
+ U^{\rm r}_{\sigma' \diagup \sigma_{i+1,j}} 
+ U^{\rm r}_{\sigma' \diagdown \sigma_{i,j-1}} 
+ U^{\rm r}_{\sigma_{i,j+1} \diagdown \sigma'} \\
- U^{\rm r}_{\sigma_{i-1,j} \diagup \sigma_{i,j}} 
- U^{\rm r}_{\sigma_{i,j} \diagup \sigma_{i+1,j}} 
- U^{\rm r}_{\sigma_{i,j} \diagdown \sigma_{i,j-1}} 
- U^{\rm r}_{\sigma_{i,j+1} \diagdown \sigma_{i,j}}.
\end{multline}
    
Once the rate of every possible reaction has been calculated, a reaction is chosen randomly. The probability of choosing reaction $r$ is
\begin{equation}
\text{Prob}(\text{selecting } r) = k_r/k_T,
\end{equation}
where $k_T$ is the sum of the rates of all possible reactions.

The system is then updated by the chosen reaction, and the system time is increased by some time $t_\text{step}$, which is exponentially distributed with mean $1/k_T$.

\section{Order parameters}\label{app:orderparams}
We use the error, density, and energy to distinguish different states in the multifarious assembly model. In this Appendix section we provide the precise definitions of these quantities.

To find the error\cite{Murugan2015PNAS}, first find the largest connected cluster of tiles $G$. Then construct $A=G \cup S^\mathrm{s}$. Here $S^\mathrm{s}$ is the full structure corresponding to the identity of the seed initially placed in the system, located at the center of the lattice. Finally calculate the overlap $O_\mathrm{s}={|A \cap S^\mathrm{s}|}/{|A|}$. The error is then
\begin{equation}
\text{Error} = 1-O_\mathrm{s},
\end{equation}
which ranges between 0 and 1.

The density is the fraction of the system filled with tiles,
\begin{equation}
\rho = \frac{1}{L^2}\sum_{\alpha} (1-\delta_{0,\sigma_\alpha}),
\end{equation}
which also ranges between 0 and 1.

The energy is defined as
\begin{equation}
E = \frac{\sum_{\langle \alpha, \beta \rangle} U^{\rm r}_{\sigma_\alpha \square \sigma_\beta}}{2L(L-1)\varepsilon},
\end{equation}
essentially counting the number of bonds in the lattice at a given point in time.
We normalize the energy with respect to the maximum system energy (fully-bonded system), such that the energy takes values between 0 and -1

\section{Calculating the rates to form and remove a strip}\label{app:KfKr}

\begin{figure}[tb]
\begin{center}
\includegraphics[width=\linewidth]{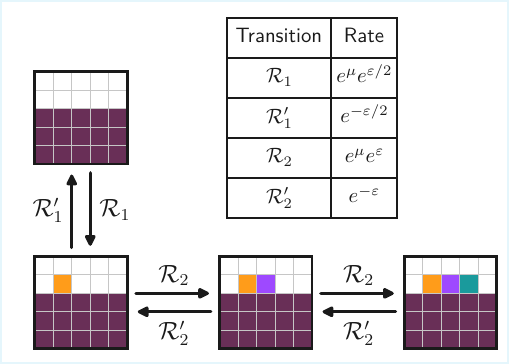}
\caption{Reactions used in the calculation of adding or removing a strip of tiles to the boundary of a structure.
\label{fig:strip_transitions}}
\end{center}
\end{figure}

We calculate the average rate to form and to remove a strip of $g$ tiles on the boundary of a structure. We use the method which is described in detail in a subsequent work\cite{Metson2025arXiv-b}. In summary, the overall rate to form a strip is the inverse of the sum of the average times for each reaction in the process. The relevant reactions and their corresponding rates are summarized in Fig.~\ref{fig:strip_transitions}.

Forming a strip of length $g$ involves one $\mathcal{R}_1$ reaction, followed by $g-1$ $\mathcal{R}_2$ reactions. The overall rate, assuming that these reactions dominate the dynamics, is
\begin{align}
K_\text{f} &= \left(\frac{1}{\mathcal{R}_1} + \frac{g-1}{\mathcal{R}_2} \right)^{-1}.
\end{align}
Similarly, the rate to remove a strip of length $g$ is
\begin{align}
K_\text{r} 
    &= \left(\frac{g-1}{\mathcal{R}'_2} + \frac{1}{\mathcal{R}'_1} \right)^{-1}.
\end{align}

The ratio of the forward and backwards rates is thus
\begin{align}
\frac{K_\text{f}}{K_\text{r}} 
    &= e^{\mu} e^{2\varepsilon} \cdot \frac{(g-1) + e^{-\varepsilon/2}}{(g-1) + e^{\varepsilon/2}},
\end{align}
where we have inserted the expressions for $\mathcal{R}_1,\mathcal{R}_1',\mathcal{R}_2$, and $\mathcal{R}_2'$.

\section{Calculating the critical seed size} \label{app:critical_seed_size}
We can use the $g$-tile strip theory to derive the critical seed size as a function of the parameters $(-\mu,\varepsilon)$. In this case we are considering the scenario where we start with a small seed of the pattern we would like the system to assemble. We know there will always be at least one valid configuration to lay a fully-bonded $g$-tile strip, namely the configuration corresponding to the next layer of the pattern. Therefore we do not need to consider the expected number of configurations, since there will always be at least one, meaning we can focus only on the energetic considerations corresponding to the ratio of the rates of laying and removing a strip of length $g$. 
From Eq.~\ref{eq:laying_favourable}, the condition for laying a strip of length $g$ becoming favorable is
\begin{equation}
    g > 1-e^{\varepsilon/2}\frac{e^{\mu+\varepsilon}-1}{e^{\mu+2\varepsilon}-1},
\end{equation}
directly giving us the critical nucleation radius
\begin{equation}\label{eq:critical_seed_radius}
    r_* = 1-e^{\varepsilon/2}\frac{e^{\mu+\varepsilon}-1}{e^{\mu+2\varepsilon}-1}.
\end{equation}
If we start with a seed which has side lengths $r>r_*$ then it is favorable to build the next layer of the pattern by adding a strip of length $r$ to the seed. Furthermore, we know that since it is not favorable to remove a strip of length $r>r_*$, the seed we start with is also stable. As the seed grows, the length of the strip we can add increases. Longer strips are energetically more favorable, and so given we start with a seed above the critical nucleation radius we will be able to assemble the full structure.
For $\varepsilon>-\mu$, we get $r_*<1$, corresponding to the fact that singly-bonded tiles are enough to lead to stable structure growth. For $-\mu/2<\varepsilon<-\mu$, we have $r_*>1$, and as $\varepsilon\to-\mu/2$ the critical radius diverges. For $\varepsilon<-\mu/2$ all structures are unstable and dissolve into dispersion.

From a classical nucleation theory perspective, we typically have a free energy of the form
\begin{equation}
    F(r) = \sigma_{\rm s} r - f_{\rm b} r^2,
\end{equation}
where $\sigma_{\rm s}$ is a surface tension and $f_{\rm b}$ is the bulk free energy density. Taking the derivative of the free energy with respect to $r$, and setting that to zero, gives the critical nucleation radius. In our case,
\begin{equation}
    \sigma_{\rm s} = e^{2\varepsilon}-1,
\end{equation}
and 
\begin{equation}
    f_{\rm b} = \frac{1}{2} \frac{\left(e^{\mu+2\varepsilon}-1\right)\left(e^{2\varepsilon}-1\right)}{e^{\mu+2\varepsilon}-1 - e^{\varepsilon/2}\left(e^{\mu+\varepsilon}-1\right)},
\end{equation}
leads to the critical nucleation radius obtained in Eq.~\ref{eq:critical_seed_radius}.
The $g$-tile strip argument presented here, which considers whether adding a new layer to the seed is energetically favorable, coincides with the free energy argument since calculating $\partial_r F$ is essentially considering the free energy change when increasing the seed radius by a small amount.
Taking the small $(\varepsilon,\mu)$ limit of our $F(r)$ leads to the free energy obtained by considering the energy of an $r\times r$ seed derived in previous work\cite{Murugan2015PNAS}
\begin{equation}
    F(r) = -2\varepsilon r(r-1) + (-\mu) r^2.
\end{equation}

\section{Calculating the expected number of chimeric strips} \label{app:CountingChimConfigs}

\begin{figure}[tb]
\begin{center}
\includegraphics[width=\linewidth]{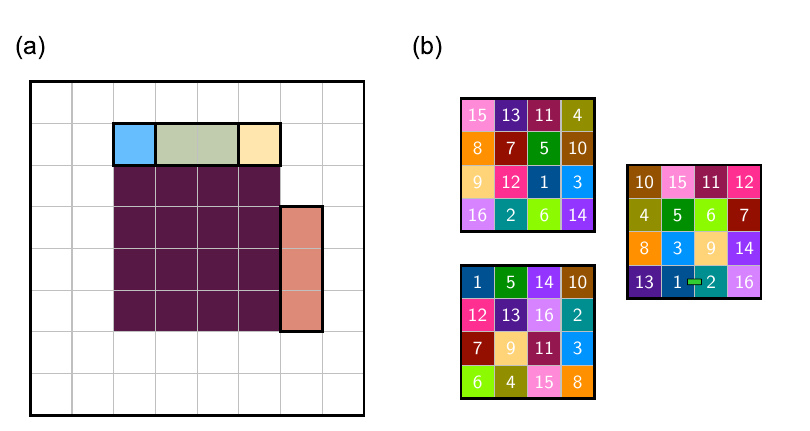}
\caption{(a) Three of the possible $4(l+1-g)=8$ locations to attach a strip of length $g=3$ to a structure of size $l=4$.
(b) Binding neighbors are determined by the random target configurations. Here tile species 1 and 2 interact in one of the structures, marked by the green bond in the structure on the right. The probability that a random pair of tiles interact in a particular direction is $\frac{m}{l^2 - 1}\left(1-\frac{1}{l}\right)$.
\label{fig:Ctilde_sketch}}
\end{center}
\end{figure}

To analyze the stability of assembled structures against chimera formation, we would like to calculate $\tilde{C}$, the number of fully-connected strips that could form on the boundary of a structure. Consider a structure of size $l\times l$ sitting in the system. For a strip of length $g$, where $g<l$, there are $4(l+1-g)$ locations for a strip of $g$ tiles to attach. This is shown in Fig.~\ref{fig:Ctilde_sketch}(a). For a given site in the strip, there are on average $(m-1)\left(1-\frac{1}{l}\right)$ suitable tiles which form a bond to the central structure. The factor $\left(1-\frac{1}{l}\right)$ corresponds to the probability that a tile is not found on a particular edge in a given structure, and is a small correction for the structure sizes we consider. For the whole strip this contributes a factor of $(m-1)^g\left(1-\frac{1}{l}\right)^g$ to $\tilde{C}$. Finally, we need to consider the $g-1$ bonds between tiles in the strip. The probability that a random pair of neighboring tiles will interact is $\min\left[1, \frac{m}{l^2 - 1}\left(1-\frac{1}{l}\right)\right]$. For the systems we consider $m\ll l^2$ and so we need not consider the $\min$ function, leading to a total contribution to $\tilde{C}$ of $\left(\frac{m}{l^2 - 1}\right)^{g-1}\left(1-\frac{1}{l}\right)^{g-1}$. The random neighbor bonds are illustrated in In Fig.~\ref{fig:Ctilde_sketch}(b). Putting all these factors together gives the full expression for $\tilde{C}$ in Eq.~\eqref{eq:number_chim_configs}.

\section{Calculating the density across the liquid-dispersion transition} \label{app:L-D_transition}
The density is the ratio of filled tiles to empty tiles in the system, which at zero interaction energy will reflect the ratio of filled tiles to empty tiles in the reservoir
\begin{align}
    \rho &= \frac{N_{\mathrm{filled}}}{N_{\mathrm{filled}} + N_{\mathrm{empty}}} \\
         &= \frac{ \frac{N_{\mathrm{filled}}}{N_{\mathrm{empty}}} }{ \frac{N_{\mathrm{filled}}}{N_{\mathrm{empty}}} + 1} \\
         &= \frac{\frac{\sum_i N_i}{N_0}}{\frac{\sum_i N_i}{N_0} + 1} \\
         &= \frac{\sum_i c_i}{\sum_i c_i + 1}.
\end{align}
In our case each of the $M$ different pattern tile species are present in the reservoir with the same concentration $c_i = e^{\mu}$, leading to
\begin{align}
    \rho &= \frac{M e^{\mu}}{M e^{\mu} + 1} \\
         &= \frac{M}{M+e^{-\mu}}. \label{eq:app:density_mu}
\end{align}

\section{Calculating the number of bonds across the chimera-liquid transition} \label{app:L-Ch_transition}
Let us define the number of bonds per tile in the system as 
\begin{equation}
\nu = \rho_1 + 2\rho_2 + 3\rho_3 + 4\rho_4,
\end{equation}
where $\rho_u$ is the density of tiles in the system making $u$ bonds. Adding a tile that increases the number of bonds in the system is favored due to the increased rate (continuous-time Gillespie) or acceptance probability (discrete-time Monte Carlo). However, configurations with more bonds are rarer due to the finite number of interactions between tile species. Using these two considerations, we write down the probability of finding a tile forming $v$ bonds in the system as
\begin{equation}
    \text{Prob}(v \text{ bonds}) = P_v e^{v\varepsilon},
\end{equation}
where $P_v$ is the probability that a tile will make $v$ bonds when added to a random site in the system. Since we are at low $|\mu|$, we will assume every lattice site is filled with a tile. Therefore, $P_v$ becomes the probability that a tile will make $v$ bonds to four randomly chosen neighboring tiles. There are $m$ structures stored in the system, so (neglecting the boundary of each structure) each tile makes a bond in a particular direction to $m$ other tile species. With $M$ tile species in total we get
\begin{align}
P_4 &= \left(\frac{m}{M}\right)^4 \\
P_3 &= 4\left(\frac{m}{M}\right)^3 \left(1-\frac{m}{M}\right) \\
P_2 &= 6\left(\frac{m}{M}\right)^2 \left(1-\frac{m}{M}\right)^2 \\
P_1 &= 4\left(\frac{m}{M}\right)^1 \left(1-\frac{m}{M}\right)^3 \\
P_0 &= \left(1-\frac{m}{M}\right)^4.
\end{align}
Using the expression for $\text{Prob}(v \text{ bonds})$ we can estimate the average number of bonds per tile as
\begin{equation}
    \nu = \frac{4P_4 e^{4\varepsilon} + 3P_3 e^{3\varepsilon} + 2P_2 e^{2\varepsilon} + P_1 e^{\varepsilon}}{Z},
\end{equation}
where the normalization factor is given by the sum of the probabilities
\begin{equation}
    Z = P_4 e^{4\varepsilon} + P_3 e^{3\varepsilon} + P_2 e^{2\varepsilon} + P_1 e^{\varepsilon} + P_0.
\end{equation}
Although not shown here, writing down time-dependent mean-field master equations for the number of tiles forming different numbers of bonds in the system also leads to this result in steady-state. Finally, using $E = -\frac{\nu}{4}$ we get to Eq.~\eqref{eq:energy_epsilon}.

In this derivation, to calculate $P_v$ we assume that the tile is being added to a random configuration of other tiles. This assumption is valid for low $\varepsilon$ in the liquid state, where the tiles are constantly churning. However it is not valid for moderate or large $\varepsilon$, leading to the observed disagreement with simulation measurements of the energy which can be seen in Fig.~\ref{fig:LChTransition}.

\end{document}